Routledge
Taylor & Francis Group

OPEN ACCESS

Check for updates

**Literature Review Corner**

# Online Behavioral Advertising: A Literature Review and Research Agenda

## Sophie C. Boerman, Sanne Kruikemeier, and Frederik J. Zuiderveen Borgesius

*University of Amsterdam, Amsterdam, the Netherlands*

Advertisers are increasingly monitoring people's online behavior and using the information collected to show people individually targeted advertisements. This phenomenon is called online behavioral advertising (OBA). Although advertisers can benefit from OBA, the practice also raises concerns about privacy. Therefore, OBA has received much attention from advertisers, consumers, policymakers, and scholars. Despite this attention, there is neither a strong definition of OBA nor a clear accumulation of empirical findings. This article defines OBA and provides an overview of the empirical findings by developing a framework that identifies and integrates all factors that can explain consumer responses toward OBA. The framework suggests that the outcomes of OBA are dependent on advertiser-controlled factors (e.g., the level of personalization) and consumer-controlled factors (e.g., knowledge and perceptions about OBA and individual characteristics). The article also overviews the theoretical positioning of OBA by placing the theories that are used to explain consumers' responses to OBA in our framework. Finally, we develop a research agenda and discuss implications for policymakers and advertisers.

In today's digital world, advertisers have seized the opportunity to use online data about consumers to personalize and target advertisements. Such data can include websites visited, articles read, and videos watched, as well as everything

Address correspondence to Sophie C. Boerman, University of Amsterdam, Amsterdam School of Communication Research, P.O. Box 15791, Amsterdam, the Netherlands. E-mail: S.C.Boerman@uva.nl

Sophie C. Boerman (PhD, University of Amsterdam) is an assistant professor of persuasive communication, Amsterdam School of Communication Research, University of Amsterdam.

Sanne Kruikemeier (PhD, University of Amsterdam) is an assistant professor of political communication, Amsterdam School of Communication Research, University of Amsterdam.

Frederik J. Zuiderveen Borgesius (PhD, University of Amsterdam) is a postdoctoral researcher, Institute for Information Law, University of Amsterdam.

searched for with a search engine. This phenomenon is called *online behavioral advertising* (OBA). In a simple example of OBA, an advertising network (i.e., a company that serves advertising on thousands of websites) tracks a consumer's website visits. If a consumer visits several websites about cars, the network assumes the consumer is interested in cars. The network can then display ads for cars only to people (presumed to be) interested in cars. Consequently, when two people visit the same website at the same time, one may see car ads while the other (who had visited websites about furniture) may see furniture ads.

Advertisers see OBA as one of the most important new ways of reaching targeted audiences. Online advertising revenues are growing rapidly and setting records every year (Interactive Advertising Bureau 2016), and it is believed that OBA will be a part of this growth (eMarketer 2010; Chen and Stallaert 2014). The industry claims that OBA creates more relevant and efficient ads and boosts ad effects (Beales 2010; Chen and Stallaert 2014). However, the practice also involves collecting, using, and sharing personal data, and thus raises consumer privacy concerns. Therefore, OBA has received much attention from regulators, such as the U.S. Federal Trade Commission (FTC 2012) and European Data Protection Authorities (Article 29 Data Protection Working Party 2010), and consumer organizations. In response, industry alliances, such as the Digital Advertising Alliance in the United States and the European Interactive Digital Advertising Alliance, set up self-regulatory programs to protect consumer privacy and describe how to inform consumers about data collection and usage.

OBA is believed to be part of the future of advertising. It is one of the new options advertisers can choose to use in their campaigns that allows for more precise targeting (Keller 2016; Kumar and Gupta 2016). Leading scholars argue that advertising will become more personalized and targeted and will involve more individual communication, where advertisers can iterate messages based on consumer behavior and needs (Kumar and Gupta 2016; Schultz 2016; Rust 2016). This emphasizes the relevance of the topic not only in practice but also in the academic field.

Despite the growing interest in OBA, a clear understanding of OBA is lacking. First, there have been various definitions of OBA, making the concept ambiguous. Second, OBA research has examined a wide range of independent, mediating, moderating, and outcome variables without a clear accumulation of knowledge. This is partly due to the interdisciplinary nature of the field and the various interested parties, including advertisers, consumers, computer scientists, and policymakers.

To address these issues, we first define OBA. Second, we provide an overview of empirical findings by developing a framework that identifies and integrates all factors that can explain consumer responses to OBA. The proposed framework provides an up-to-date review of this new, and still developing, type of advertising. In addition, the framework identifies the most important factors in examining, predicting, and evaluating the outcomes of OBA, making the framework relevant not only to academic research but also to advertisers. Third, we establish the theoretical positioning of OBA by reviewing theories used to explain people's responses in the context of our framework. Finally, our framework and the theoretical positioning of OBA help identify gaps in the literature and facilitate the development of a research agenda.

## DEFINING ONLINE BEHAVIORAL ADVERTISING

There are many definitions of OBA, which is also called "online profiling" and "behavioral targeting" (Bennett 2011). Examples include "adjusting advertisements to previous online surfing behavior" (Smit, Van Noort, and Voorveld 2014, p. 15), "a technology-driven advertising personalization method that enables advertisers to deliver highly relevant ad messages to individuals" (Ham and Nelson 2016, p. 690), and "the practice of collecting data about an individual's online activities for use in selecting which advertisement to display" (McDonald and Cranor 2010, p. 2). These definitions and others have two common features: (1) the monitoring or tracking of consumers' online behavior and (2) use of the collected data to individually target ads. Therefore, we define OBA as *the practice of monitoring people's online behavior and using the collected information to show people individually targeted advertisements*. Online behavior can include web browsing data, search histories, media consumption data (e.g., videos watched), app use data, purchases, click-through responses to ads, and communication content, such as what people write in e-mails (e.g., via Gmail) or post on social networking sites (Zuiderveen Borgesius 2015a).

OBA could be considered a type of personalized or customized advertising—concepts which refer to tailoring advertising to individuals. However, these concepts have a broader scope than OBA and could include advertising amended to personal data that are not based on online behavior. OBA refers only to advertising that is based on people's online behavior.

To track consumers' browsing behavior, companies often use (tracking) cookies, but other technologies include flash cookies and device fingerprints (Altaweel, Good, and Hoofnagle 2015). Recently, researchers found that the 100 most popular sites collect more than 6,000 cookies, of which 83% are third-party cookies, with some individual websites collecting more than 350 cookies (Altaweel, Good, and Hoofnagle 2015). These cookies allow companies to collect detailed information about millions of consumers, partly for use in OBA. To illustrate the magnitude of this business, Facebook has individual profiles of 1.65 billion people (Facebook 2016) and AddThis has profiles of 1.9 billion people (AddThis 2016).

OBA differs from other types of online advertising because it aims at personal relevance, which often happens covertly. Similar to other forms of personalized advertising, such as location-based advertising (e.g., Ketelaar et al. 2017) and ads that include people's names (Bang and Wojdynski 2016), OBA uses personal information to tailor ads in such a way that they are perceived as more personally relevant. A new dimension to this personalization is the fact that the tracking of online activities, collection of behavioral data, and dissemination of information often happen covertly (Ham and Nelson 2016; Nill and Aalberts 2014). This covertness may be harmful and unethical, as consumers are unaware of the persuasion mechanisms that entail OBA; it has led to a call for transparency.

## LITERATURE SEARCH

We performed a keyword search of the most important electronic databases in advertising and communication science (i.e., PsycINFO, Web of Science, Communication and Mass Media Complete, Academic Search Premier, database of the World Advertising Research Center). The keywords used were "online behavioral/behavioural advertising," "online behavioral/behavioural targeting," "customized/customised advertising," "personalized/personalized (online) advertising," and "online profiling." The search period covered all manuscripts available by the end of September 2016. After identifying a study, we examined its references to find further studies. In addition, we contacted several experts in the field to inquire about other relevant manuscripts.

We included studies that reported on empirical data while leaving out nonempirical studies, such as legal studies. Furthermore, by definition, OBA involves tailoring advertising based on online behavior. Therefore, we excluded studies that addressed personalized advertising based on personal data that were not inferred from online behavior. In total, 32 manuscripts fit the criteria, and the earliest study was published in 2008. Among these, 21 were from academic journals, six from conference proceedings, one was a book chapter, and four were published online.

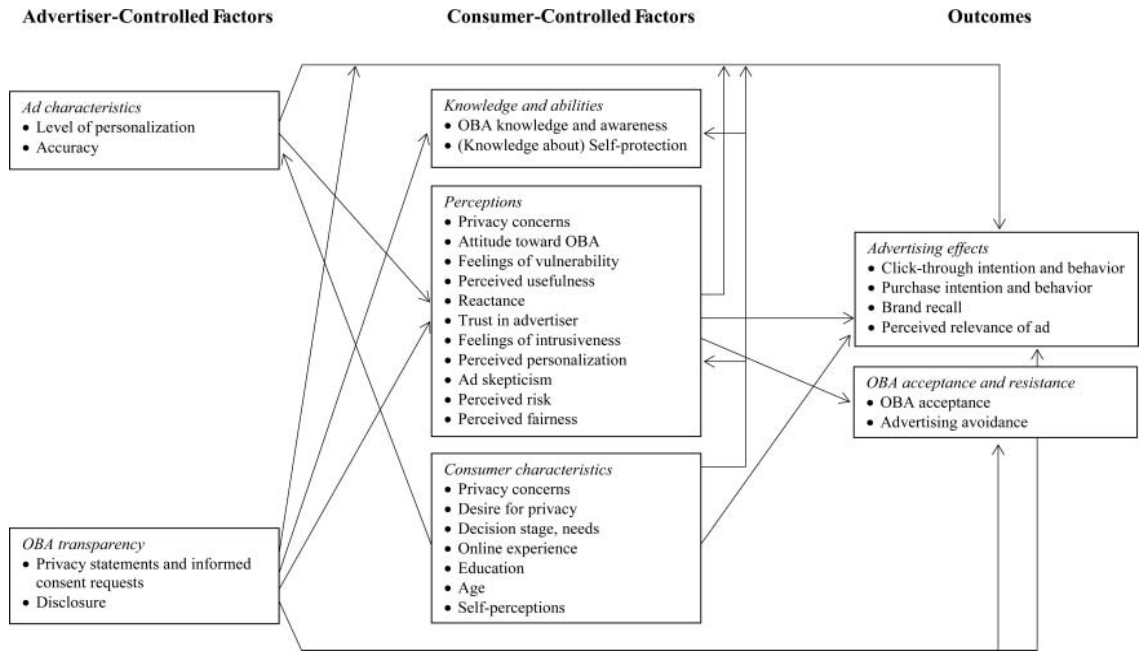

FIG. 1.   Proposed framework of online behavioral advertising (OBA).

Based on these studies, we developed a framework that identifies the factors that explain consumer responses to OBA and illustrates their interconnectedness (see Figure 1). The framework focuses on relationships with empirical support. To develop the framework, we first identified all variables that were studied with regard to OBA and grouped them into three main factors based on the interactive advertising model (Rodgers and Thorson 2000). This model explains how consumers perceive and process online ads and distinguishes three main types of factors: advertiser-controlled factors, consumer-controlled factors, and advertising outcomes. Within these factors, we distinguished separate aspects. The advertiser-controlled factors include (a) the ad characteristics, or the factors which are part of the ad itself and which can differ among different online behavioral ads, and (b) the forms of transparency that advertisers use to communicate that an ad is based on online behavior. These forms of transparency involve information that often accompanies an ad (e.g., an additional logo or privacy statement), which some self-regulatory bodies require for OBA.

The consumer-controlled factors include (a) a cognitive aspect, including people's knowledge and abilities with respect to OBA; (b) an affective aspect, including people's perceptions of OBA in general or of a specific ad; and (c) personal characteristics, such as a person's age or desire for privacy. Finally, the outcomes include consumers' responses to OBA with respect to (a) the actual advertising effects, such as purchases and click-through rates, and (b) the degree to which people accept or avoid OBA. Table 1 provides an overview of the literature addressing these factors.

## ADVERTISER-CONTROLLED FACTORS

### Ad Characteristics

*Level of personalization.*  The data used to create OBA vary widely. Because advertisers typically do not use all these data for one ad, the levels of personalization differ. We propose that the level of personalization is based on (a) the types of personal data that are used to target the ad (e.g., browsing data or search history) and (b) the amount of information that is used (e.g., just one search term or a combination of browsing data and search history). Several studies compared different levels of personalization in OBA. The types of information used included age, gender, location (Aguirre et al. 2015), education level (Tucker 2014), interests (Aguirre et al. 2015; Tucker 2014), online shopping behavior (Bleier and Eisenbeiss 2015), and search history (Van Doorn and Hoekstra 2013). The researchers created various levels of personalization by combining one or more types of information. Their findings suggest that the level of personalization influences consumer-related factors, such as feelings of intrusiveness (Van Doorn and Hoekstra 2013), feelings of vulnerability (Aguirre et al. 2015), the ad's perceived usefulness, reactance, and privacy concerns (Bleier and Eisenbeiss 2015). The level of personalization also influences OBA outcomes, such as click-through intentions and rates (Aguirre et al. 2015; Bleier and Eisenbeiss 2015; Tucker 2014).

Negative responses to higher levels of personalization correspond to choice theory, psychological ownership theory, and psychological reactance theory (Aguirre et al. 2015; Bleier and Eisenbeiss 2015; Tucker 2014). These theories propose that people want to restore their freedom of choice, control,

TABLE 1
Summary of Framework-Related Literature

| Factors | References |
| --- | --- |
| **Advertiser-controlled factors** | |
| *Ad characteristics* | |
| Level of personalization | Aguirre et al. 2015; Bleier and Eisenbeiss 2015; Tucker 2014; Van Doorn and Hoekstra 2013 |
| Accuracy | Summers, Smith, and Reczek 2016 |
| *OBA transparency* | |
| Privacy statements and informed consent requests | Marreiros et al. 2015 |
| Disclosure | Aguirre et al. 2015; Jai, Burns, and King 2013; Leon, Cranshaw, et al. 2012; Miyazaki 2008; Schumann, von Wangenheim, and Groene 2014; Ur et al. 2012; Van Noort, Smit, and Voorveld 2013 |
| **Consumer-controlled factors** | |
| *Knowledge and abilities* | |
| OBA knowledge | Marreiros et al. 2015; McDonald and Cranor 2010; Smit, Van Noort, and Voorveld 2014; Ur et al. 2012; Ham and Nelson 2016 |
| (Knowledge about) self-protection | Balebako et al. 2012; Cranor 2012; Leon, Ur, et al. 2012 |
| *Perceptions*[*] | |
| Privacy concerns (dv, med) | Aguirre et al. 2015; Antón, Earp, and Young 2010; Baek and Morimoto 2012; Bleier and Eisenbeiss 2015; Ham and Nelson 2016; Jai, Burns, and King 2013; Lambrecht and Tucker 2013; Lee et al. 2015; Marreiros et al. 2015; McDonald and Cranor 2010; Moore et al. 2015; Phelan, Lampe, and Resnick 2016; Schaub et al. 2016; Stanaland, Lwin, and Miyazaki 2011; Smit, Van Noort, and Voorveld 2014; Sutanto et al. 2013; Turow et al. 2009; Turow, Carpini, and Draper 2012; Ur et al. 2012; Van Doorn and Hoekstra 2013; Yang 2013 |
| Attitude toward OBA | |
| Feelings of vulnerability (med) | |
| Perceived usefulness (med) | |
| Reactance (med) | |
| Trust in advertiser (dv, mod) | |
| Feelings of intrusiveness (med) | |
| Perceived personalization (iv) | |
| Ad skepticism (iv) | |
| Perceived risk (med) | |
| Perceived fairness (med) | |
| *Consumer characteristics* | |
| Privacy concerns | Antón, Earp, and Young 2010; Baek and Morimoto 2012; Lee et al. 2015; Smit, Van Noort, and Voorveld 2014; Yang 2013 |
| Desire for privacy | Miyazaki 2008; Stanaland, Lwin, and Miyazaki 2011 |
| Decision stage, needs | Lambrecht and Tucker 2013; Van Doorn and Hoekstra 2013 |
| Online experience | Lee et al. 2015; Miyazaki 2008 |
| Education | Smit, Van Noort, and Voorveld 2014 |
| Age | Smit, Van Noort, and Voorveld 2014; Turow et al. 2009 |
| Self-perceptions | Summers, Smith, and Reczek 2016 |
| **Outcomes** | |
| *Advertising effects* | |
| Click-through intention and behavior | Lambrecht and Tucker 2013; Van Doorn and Hoekstra 2013; Summers, Smith, and Reczek 2016 |
| Purchase intention and behavior | Aguirre et al. 2015; Bleier and Eisenbeiss 2015; Tucker 2014 |
| Brand recall | Van Noort, Smit, and Voorveld 2013 |
| Perceived relevance of ad | Van Noort, Smit, and Voorveld 2013 |
| *OBA acceptance and resistance* | |
| OBA acceptance | Schumann, von Wangenheim, and Groene 2014 |
| Advertising avoidance | Baek and Morimoto 2012 |

*Note.* *The abbreviations between parentheses clarify the role of this factor in the studies (if applicable): dv = dependent variable, iv = independent variable, med = mediator, mod = moderator. To avoid repetition, the References cell sums up all articles that studied one or more of the perceptions.

and ownership when they feel they are threatened. Highly personalized ads lead people to perceive a loss of choice, control, or ownership, and thus cause negative feelings and responses.

*Accuracy.* Another key characteristic of OBA is accuracy. Based on self-perception theory, Summers, Smith, and Reczek (2016) propose that OBA can act as an implied social label. When consumers know an ad is based on their past online behavior, they understand that the marketer has made inferences about them. Thus, OBA provides an external characterization of the self, leading consumers to adjust their self-perceptions and draw on these perceptions to determine their purchase behavior (Summers, Smith, and Reczek 2016). Interestingly, these effects seem to occur only when OBA is accurately connected to past behavior.

## OBA Transparency

Research and regulations also indicate the importance of transparency about OBA. Consumers want openness and to be informed about the collection, usage, and sharing of personal data (Gomez, Pinnick, and Soltani 2009; Turow et al. 2009). Privacy laws require companies to be transparent about their data processing practices (EU Data Protection Directive 1995; EU General Data Protection Regulation 2016). Similarly, the U.S. FTC (2012) calls for transparency regarding OBA.

*Privacy Statements and Informed Consent Requests.* Companies typically publish privacy statements on their websites to comply with transparency requirements. In addition, some laws aim to improve transparency by requiring companies to obtain consent before using OBA. The FTC (2012) also emphasizes that companies should offer consumers choices regarding OBA. According to the privacy guidelines from the Organisation for Economic Co-Operation and Development (OECD 2013), personal data should be obtained where appropriate, with the knowledge or consent of the data subject. One of the goals in these efforts is empowering the consumer. Privacy laws typically aim to enable consumers to make informed decisions about privacy and personal data. For instance, some consumers might like OBA and thus allow companies to track them, while others may prefer more privacy and decline tracking.

A privacy statement is a document on a website that discloses which personal data are collected through the website, as well as how and why. In theory, privacy statements should help reduce the information asymmetry between companies and consumers through companies disclosing information to consumers (McDonald and Cranor 2008). Although inclusion of privacy statements disclosing the usage of cookies has increased (Miyazaki 2008), such statements are seldom read and thus fail to inform consumers (Cranor 2003; McDonald and Cranor 2008; Milne and Culnan 2004). It would take a person approximately 201 hours per year to read all privacy statements for the websites he or she visits (McDonald and Cranor 2008); in addition, people are highly unlikely to understand the language of such statements (Jensen and Potts 2004; Milne, Culnan, and Greene 2006). People tend to agree with almost all requests, or they simply ignore them (Marreiros et al. 2015; Zuiderveen Borgesius 2015b). Thus, informed consent requests seem to be a valuable approach to give people control, yet they fail to inform or empower people.

*Disclosure.* The online marketing industry has developed self-regulatory approaches to improve transparency that entail explicit disclosure of data collection, usage, and distribution. The current disclosure methods used by the industry involve icons, logos, and taglines. The Digital Advertising Alliance in the United States and the European Interactive Digital Advertising Alliance have developed a standard icon that consists of the letter *i* in a blue triangle. However, advertisers can use other form of disclosure, such as pop-ups or text that explain OBA.

Different academic studies have compared the effectiveness of OBA icons. Leon, Cranshaw, et al. (2012) found that disclosures are rarely noticed: Only one-quarter of the respondents remembered OBA disclosure icons (the standard icon and an "asterisk man" icon), and only 12% remembered seeing a tagline (e.g., "Why did I get this ad?" or "AdChoices") and correctly selected the tagline they had seen from a list. In addition, none of the taglines were understood to be links to pages where you can make choices about OBA, nor did they increase knowledge about OBA. Other studies show that consumers are unfamiliar with the icons (Ur et al. 2012; Van Noort, Smit, and Voorveld 2013), do not understand their purpose (Ur et al. 2012), and rarely notice them (Van Noort, Smit, and Voorveld 2013). However, the standard icon could effectively increase OBA awareness and understanding when accompanied by an explanatory label stating, "This ad is based on your surfing behavior" (Van Noort, Smit, and Voorveld 2013).

Consumers do seem to appreciate companies' transparency initiatives (Van Noort, Smit, and Voorveld 2013). When firms do not openly state that they use personal data to personalize ads and then present highly personalized ads, consumers feel more vulnerable (Aguirre et al. 2015). However, when companies are open about data collection, it does not affect perceived vulnerability. In addition, Miyazaki (2008) found that explicitly disclosing the usage of cookies in a privacy statement and a pop-up can increase consumers' trust toward the website and their intentions to use and recommend it. Hence, advertisers benefit from transparency about OBA.

The effects of transparency on consumer trust can be explained by social contract theory (Miyazaki 2008) and expectancy violation theory (Moore et al. 2015). According to social contract theory, advertisers form an implied social contract with consumers by explicitly disclosing the collection and use of personal information. Under such an implied contract, consumers expect advertisers to collect and care for their personal information in a responsible manner. When companies do not disclose the collection and use of personal information, or do not use the information responsibly, they are violating this contract. In addition, expectancy violation theory (see Moore et al. 2015) suggests that a violation of personal space will drive subsequent reactions. Thus, when an advertiser collects and uses information without disclosing it and without consent, this may lead to a violation of the social contract, a violation of personal space, and, as a result, lowers trust.

## CONSUMER-CONTROLLED FACTORS

*OBA Knowledge and Awareness.* Several academic studies show that consumers have little knowledge about OBA and hold misconceptions (Marreiros et al. 2015; McDonald and Cranor 2010; Smit, Van Noort, and Voorveld 2014). Moreover, people have little insight into the extent to which their online behavior is tracked (Ur et al. 2012). Interestingly, the perception of having knowledge about OBA makes people more likely to perceive the effects of OBA as larger on others than on themselves (Ham and Nelson 2016). Even more confusion arises in regard to legal protections: A substantial majority of Americans have false beliefs about companies' rights to share and sell information about them online (McDonald and Cranor 2010; Turow et al. 2009). These findings suggest there is information asymmetry: Companies know much about consumers, yet consumers know little about what happens to their personal data. It seems nearly impossible for people to determine which companies collect which personal data online and what happens to the data.

These findings indicate that consumers' mental models (i.e., their beliefs about how a system works, interacts, or behaves; see McDonald and Cranor 2010) and persuasion knowledge (i.e., consumers' knowledge and beliefs about persuasive tactics; see Baek and Morimoto 2012; Van Noort, Smit, and Voorveld 2013; Ham and Nelson 2016) are rarely well developed in the context of OBA. In addition, there seems to be an important relationship between persuasion knowledge and people's third-person perceptions about OBA. The more people think they know about how OBA works (i.e., subjective persuasion knowledge), the more they tend to overestimate the effects of OBA on others and underestimate its effects on themselves (Ham and Nelson 2016). This can be a problem because incorrect mental models, low persuasion knowledge, and an underestimation of OBA effects may undermine careful and educated decision making.

*(Knowledge about) Self-Protection.* Consumers' lack of knowledge about OBA impedes them from having control over their personal data (Cranor 2012). This is especially interesting because consumers want to have control over the collection and use of their personal data (Gomez, Pinnick, and Soltani 2009; Turow, Carpini, and Draper 2012). Research has shown that a minority of consumers do try to control their personal data by deleting cookies, by not letting cookies save to the hard drive, and by using software that deletes cookies. However, despite taking such actions, it appears that people do not know *why* they do this (McDonald and Cranor 2010). Protective behavior does seem to depend on consumer characteristics: the more concerned people are about their privacy, the more they try to protect their online privacy (Smit, Van Noort, and Voorveld 2014).

Moreover, not all available tools and tactics are effective for protecting privacy. Research has shown that certain tools that block third-party cookies can effectively limit OBA (Balebako et al. 2012). Opt-out options limit receiving behaviorally targeted ads but may not limit being tracked. The "do-not-track" option in browsers limits OBA only slightly. Furthermore, consumers do not seem to understand the available tools and thus have difficulties protecting their online privacy (Cranor 2012; Leon, Ur, et al. 2012).

## Consumer Perceptions

Many empirical studies on OBA consider consumer perceptions either as a consumer characteristic in a survey or as moderator, mediator, or dependent variable in experimental research. Overall, consumers' perceptions of OBA appear to be mixed. Some see the benefits of targeted ads (McDonald and Cranor 2010; Ur et al. 2012), but the majority seem to be skeptical toward OBA and find it invasive and "creepy" (Smit, Van Noort, and Voorveld 2014; Ur et al. 2012). Most adults in the United States do not want advertising to be tailored to their personal information (Turow, Carpini, and Draper 2012). Consumers believe that invasive tactics, such as using and gathering personal data, tracking, and invading a consumer's personal space, can be considered "creepy marketing" (Moore et al. 2015). These negative perceptions and privacy concerns can be explained by social presence theory. Social presence describes the feeling of being with another in mediated communication (see Phelan, Lampe, and Resnick 2016). When a computer collects your data, it generates the same negative feelings as when another person looks over your shoulder as you browse (Phelan, Lampe, and Resnick 2016).

Over the past decade, consumers have become more concerned about OBA practices and especially about their privacy (e.g., Antón, Earp, and Young 2010; McDonald and Cranor 2010). Researchers found signs for a chilling effect: People report that they change their online behavior when they know their data are being collected (McDonald and Cranor 2010). Perceptions of OBA also depend on consumer characteristics such as age. Younger people are less likely to oppose OBA compared to older people, although the majority of young people do not want OBA (Turow et al. 2009).

The notion of privacy calculus is often used to describe the process in which consumers assess the benefits and risks of OBA (Jai, Burns, and King 2013; Phelan, Lampe, and Resnick 2016; Schumann, von Wangenheim, and Groene 2014). Privacy calculus is rooted in theories such as social exchange theory (Schumann, von Wangenheim, and Groene 2014) and the acquisition-transaction utility theory (Baek and Morimoto 2012). The social exchange theory stems from psychology and proposes that people evaluate social exchanges in terms of costs and rewards. People are supposed to alter their behavior according to their evaluation and are expected to participate in social exchanges only when the rewards outweigh the costs (Schumann, von Wangenheim, and Groene 2014). Acquisition-transaction utility theory is often used to understand ethical issues in marketing and suggests that the probability of consumers purchasing a product or service depends on the perceived benefits compared with the perceived costs (Baek and Morimoto 2012). Based on these theories, consumers should accept OBA only if the benefits (e.g., personal relevance) outweigh the costs or risks (e.g., privacy invasions).

The information boundary theory (Sutanto et al. 2013) provides some insight into the actual weight that consumers give to the benefits and risks of OBA. This theory suggests people find the collection and use of personal information intrusive and thus perceive it as a risk or cost that does not outweigh the possible benefits of OBA when people consider using this information as crossing a boundary. When a person considers the collected information as harmful or too uncomfortable, the costs do not outweigh the benefits of OBA.

In line with these theories, research has shown that privacy concerns and trust play important roles in consumer acceptance and the effectiveness of OBA. For instance, more trusted retailers can increase the perceived usefulness of their ads by developing ads that reflect consumers' interests in a complete way (Bleier and Eisenbeiss 2015). Trust can be enhanced by including a "privacy trustmark" (i.e., a symbol explaining that the website is involved in a program that protects consumers privacy), which positively affects consumers' perceptions of the trustworthiness of the advertiser, lowers privacy concerns about the advertiser, and leads to more positive behavior intentions (Stanaland, Lwin, and Miyazaki 2011). Furthermore, Sutanto and colleagues (2013) note that if consumers' personal data are not transmitted to third parties, consumers are less concerned about their privacy and more satisfied with the content of a smartphone app about products.

## Consumer Characteristics

Responses to OBA also differ among consumers, and individual levels of privacy concerns are especially important. Multiple studies suggest that people with low levels of privacy concerns or less desire for privacy tend to be more positive toward OBA (e.g., Baek and Morimoto 2012; Miyazaki 2008; Smit, Van Noort, and Voorveld 2014; Stanaland, Lwin, and Miyazaki 2011). Moreover, the level of privacy concern appears to moderate the effects of OBA on consumers' advertising responses (e.g., Lee et al. 2015; Miyazaki 2008).

The effects of OBA on purchase intentions and behavior are more positive when the ad fits consumers' needs (Van Doorn and Hoekstra 2013) and when consumers have narrowly construed preferences (construal level theory; Lambrecht and Tucker 2013). Furthermore, responses to OBA seem to be related to age, education, and online experience (Lee et al. 2015; Miyazaki 2008; Smit, Van Noort, and Voorveld 2014; Turow et al. 2009).

## OBA OUTCOMES

Several academic studies show that the outcomes of OBA are determined by factors controlled by the advertisers and by the consumer. Outcome measures have included advertising effects (i.e., click-through intentions and rates, actual purchases, and purchase intentions) and measures of OBA acceptance and avoidance. Overall, the findings are much more nuanced than the industry's promise that OBA boosts ad effects.

## Advertising Effects

### Click-through intention and click-through rates

Several studies demonstrated that the level of personalization in OBA influences click-through intentions and click-through rates. Tucker (2014) found that Facebook ads targeting a person's interests (e.g., a celebrity of whom a person is a fan) led to higher click-through rates than ads targeting background characteristics (i.e., the college a person is attending). In addition, Aguirre et al. (2015) showed that moderately personalized Facebook ads (based on interest in a subject) increased click-through rates compared to nonpersonalized ads, whereas highly personalized ads (based on interest in a subject, age, gender, and location) decreased click-through rates. Bleier and Eisenbeiss (2015) found that highly personalized banner ads (showing items consumers placed in their virtual shopping cart during a recent shopping session) increased click-through rates compared to ads with a lower level of personalization (showing items that consumers viewed during a shopping session). This effect occurred only when the ad concerned a trusted retailer.

Furthermore, several studies indicate that transparency and consumer awareness of OBA alter consumers' responses to online behavioral ads. For instance, when companies overtly inform people about the collection and use of data to personalize ads, OBA increases click-through rates. However, when companies covertly collect information, click-through rates are not influenced or are even reduced (Aguirre et al. 2015). Moreover, the difference in click-through intentions between overt and covert ads disappeared when an OBA icon was included (Aguirre et al. 2015). Likewise, an OBA icon improves brand recall and the perceived relevance of the advertised brand and the online ad (Van Noort, Smit, and Voorveld 2013). Hence, people seem to appreciate company transparency, and an icon can function as a cue to trust the advertiser and even positively affect advertising outcomes.

Thus, this self-regulatory approach does benefit advertisers but does not help consumers make informed decisions.

Research also specifies various moderators and mediators that influence the effects of OBA on click-through rates and intentions. Aguirre et al. (2015) demonstrated that the effects of OBA were mediated by the consumers' experience of vulnerability. Overall, personalization appears to increase click-through rates and intentions, but only when consumers know that data are collected. When consumers are unaware that data are collected, they feel more vulnerable when confronted with personalized advertising, which decreases their intention to click on an ad (Aguirre et al. 2015).

This idea of vulnerability is in line with the finding that trust in a retailer is a major determinant, and that the perceived usefulness of the ad, the reactance, and privacy concerns are important underlying mechanisms for the effects of OBA on click-through intentions (Bleier and Eisenbeiss 2015). For banner ads from trusted retailers, click-through intentions appear particularly high when ads reflect a combination of high depth (using items previously placed in an online shopping cart) and narrow breadth (only one out of three items shown in the ad personalized) of personalization. People perceive such ads to be more useful and do not elicit increased reactance or privacy concerns compared to low-depth ads. For ads from less-trusted retailers, a higher depth of personalization decreased click-through intentions, irrespective of the breadth of personalization.

The lower click-through intention can be explained by people experiencing lower usefulness, more reactance, and more privacy concerns. These findings are in line with the stimulus-organism-response model, which posits that stimuli influence individuals' cognitive and affective responses, which then translate into specific behavior (see Bleier and Eisenbeiss 2015; Jai, Burns, and King 2013). Indeed, OBA seems to first trigger affective responses, such as feelings of vulnerability, reactance, and privacy concerns, which consequently affect behavior (e.g., lower intentions to click on an ad).

*Purchases and purchase intention*

Research has also examined the effects of OBA on purchase intentions and actual purchases. Lambrecht and Tucker (2013) compared the effects on purchase intentions and actual purchases between OBA showing an image of previously browsed hotels and random generic ads for a travel firm. They found that the effects of the different ads depended on the decision stage of the consumer, with OBA being more effective when consumers had narrowly construed preferences and thus had a greater focus on specific and detailed information. When preferences were still broad and people were in the early stages of a purchase decision, generic ads led to a higher likelihood of purchases (Lambrecht and Tucker 2013).

As for click-through behavior, there also seem be important mediating variables that influence purchase behavior in response to OBA, which is in accordance with the stimulus-organism-response model. Van Doorn and Hoekstra (2013) conclude that more personalization increases feelings of intrusiveness and thus negatively affects purchase intentions. These negative effects cannot be offset by offering a discount, but they can be partly mitigated by ensuring the ad fits the consumer's current needs. However, an ad with high fit also increases the perceived intrusiveness of the ad. Moreover, although privacy concerns did not alter the effects of OBA, the study indicated that people with high concerns had lower intentions to purchase.

Furthermore, Summers, Smith, and Reczek (2016) emphasize the importance of consumers' self-perceptions as a mediator of the effects of OBA on purchase intentions. They found that OBA can influence consumers' self-perceptions as it reflects their past online behavior. When the ad accurately fits their behavior and perceptions, it can increase their purchase behavior.

## OBA Acceptance and Resistance

In addition to these advertising effects, empirical research has also examined which factors can explain OBA acceptance and avoidance. Baek and Morimoto (2012) found that privacy concerns and ad irritation both increase ad skepticism, which consequently leads to more avoidance of OBA. In addition, the more consumers feel the ad is personalized for them (i.e., perceived personalization), the less they avoid the ad.

Transparency about the reason why a company collects data also influences responses to ads. Schumann, von Wangenheim, and Groene (2014) compared people's responses to two arguments for data collection: a reciprocity argument which argues that the service on the website is for free in return for personal data, or a relevance argument stating data collection is necessary to make advertisements more personally relevant. The researchers found that the reciprocity argument increased consumers' acceptance of OBA compared to the relevance argument, and consumers were more likely to opt in and disclose personal data for OBA purposes. This finding suggests that consumers believe that receiving web services for "free" in return for use of their personal data is an acceptable trade-off.

People's responses to data collection depend on what happens to the data. Jai, Burns, and King (2013) showed that telling respondents that the website shared their personal and website navigation data with third-party companies increased the perceived risk and unfairness, leading to lower repurchase intentions compared to telling them their data would be shared only internally within the corporate family.

## THEORETICAL POSITIONING OF OBA

We reviewed the theories used to study OBA to gain an understanding of its theoretical positioning. Figure 2 shows an overview of theories used to explain the effects of different factors in our framework and the responses to them. This model shows that the theoretical background of the research regarding OBA is very fragmented. Almost half of the studies ($n = 15$) did not refer to specific theories. The studies that do use them employ a variety of theories from different disciplines, such as social and cognitive psychology, communication, and marketing. There is certainly no single overarching or reoccurring theory that is used to describe and explain responses to OBA.

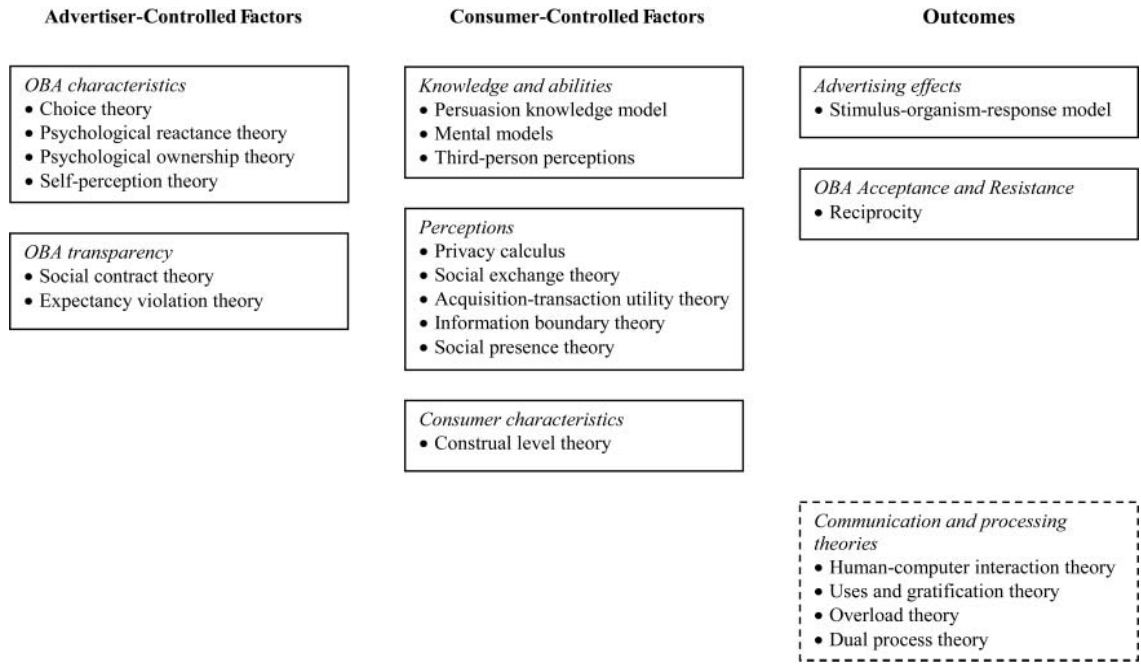

FIG. 2. Overview of theoretical position of online behavioral advertising (OBA).

The theories cited more than twice are the persuasion knowledge model (Baek and Morimoto 2012; Van Noort, Smit, and Voorveld 2013; Ham and Nelson 2016), psychological reactance theory (Aguirre et al. 2015; Baek and Morimoto 2012; Bleier and Eisenbeiss 2015; Tucker 2014), and social contract theory (Jai, Burns, and King 2013; Miyazaki 2008; Yang 2013). These three theories are all used to explain why people may have negative feelings toward OBA and resist it. When people develop persuasion knowledge about tactics used in OBA, they can become more skeptical of them (Baek and Morimoto 2012). In addition, highly personalized ads can threaten consumers' perceived ability to avoid being closely observed by firms, causing reactance (Bleier and Eisenbeiss 2015). And when people feel that advertisers violate a social contract by wrongly collecting and using their information, it can also cause distrust (Miyazaki 2008).

Although privacy concerns are an important aspect of OBA, many ($n = 13$) of the studies that discuss people's privacy concerns never relate these concerns to any kind of theoretical notions. When they do, privacy concerns are often placed within a debate about the privacy calculus, which is directly related to social exchange theory (Schumann, von

Wangenheim, and Groene 2014), acquisition-transaction utility theory (Baek and Morimoto 2012), and information boundary theory (Sutanto et al. 2013).

Aside from the theories that can be attributed to one of the factors in our framework, some general communication and processing theories are also applied to OBA: human-interaction theory (Ahrens and Coyle 2011), uses and gratifications theory (Sutanto et al. 2013), overload theory (Schumann, von Wangenheim, and Groene 2014), and dual process theory (Phelan, Lampe, and Resnick 2016). These theories are helpful in explaining how people process and respond to computer-mediated communication and specific messages, but they do not seem to be specific to the concept of OBA or our framework.

## RESEARCH AGENDA

### Theoretical Advancement

Our review of the theoretical positioning of empirical research regarding OBA demonstrates that the field is fragmented and lacks a solid theoretical basis. To advance the literature on OBA, it is important to develop more conceptual coherence between the different theoretical ideas that focus on the roles of the advertiser and consumer variables in explaining consumer responses. Valkenburg and Peter (2013) proposed that connecting different conceptual approaches could help shed light on media effects. Moreover, they suggested that research should focus on specific models that combine moderating and mediating processes (Valkenburg and Peter 2013). Thus far, most OBA studies have focused on one perspective and singled out specific moderators and mediators. However, the combinations of certain moderating and mediating processes are expected to play crucial roles.

Therefore, we urge researchers to take a more integrative look, combine insights from different perspectives, and look at the reinforcing role of outcomes in affecting future use or exposure to ads. For instance, choice theory, psychological ownership theory, and psychological reactance theory (Aguirre et al. 2015; Bleier and Eisenbeiss 2015; Tucker 2014) suggest that people perceive advertising as invasive when an ad is too personal (which is also a boundary condition as it depends on the characteristics of the ad). When people perceive a lack of freedom of choice, it could lead to heightened irritation and skepticism toward OBA, as well as more concerns about privacy (Baek and Morimoto 2012). This might ultimately contribute to a reinforcing process in which responses to OBA also affect future media use for shorter or longer periods of time (transactional media effect; Valkenburg and Peter 2013). Such reinforcing processes are currently neglected in the literature, which mainly focuses on immediate responses while ignoring future consequences.

In addition, one of the key features of OBA is that it is often covert. The perceived costs and benefits of OBA (e.g., the privacy calculus) may therefore depend on peoples' knowledge about the practice of OBA. Those with more knowledge might perceive the cost and benefits differently and might believe that the negative consequences are more severe. Our framework reveals an absence of research on the roles of consumer knowledge and abilities in OBA perceptions and responses. Combinations of theories, such as the privacy calculus notion and the persuasion knowledge model, could be useful for attaining deeper understanding of OBA transparency, as well as the antecedents and effects of consumer knowledge and abilities.

Other theories could also be added to the theoretical positioning of consumer reactions to OBA. For instance, the notion of present bias describes tendencies to choose immediate gratification and disregard future costs or disadvantages (Acquisti and Grossklags 2007; O'Donoghue and Rabin 1999). Online, people often want to read an article, watch a video, or purchase a product immediately, thus skipping privacy statements and accepting informed consent requests without thinking about long-term consequences. Despite its relevance, there is a lack of research that looks at whether the idea of present bias can explain people's behavior in response to OBA. The circumstances in which immediate gratification is sought and the motivations that play a role warrant investigation. Such motivations function as boundary conditions. For instance, when people need information immediately, certain privacy concerns might be overruled, but perhaps only when the website is trusted. Thus, we believe that the field could benefit not only from a more integrative perspective but also from the inclusion of such theories.

## Understanding Acceptance of and Resistance Toward OBA

Prior empirical research shows that consumers avoid and dislike some ads but accept others. However, it is currently not well understood why people reject OBA (Turow et al. 2009). Consumers' most important concerns seem to be privacy and a lack of control over personal data (McDonald and Cranor 2010; Smit, Van Noort, and Voorveld 2014; Turow et al. 2009; Ur et al. 2012). People might have a "general antagonism to being followed without knowing exactly how or with what effects" (Turow et al. 2009, p. 4). For instance, Yang (2013) found that consumers who are more concerned about privacy are less likely to trust online companies to protect their privacy. In addition, people with low levels of privacy concerns who are willing to share information respond more positively to OBA (Lee et al. 2015). However, there appears to be a privacy paradox: People say they care about privacy and do not intend to share personal data, but in reality they provide their data in exchange for small benefits or conveniences (Norberg, Horne, and Horne 2007). In other words, although people say they reject OBA, they take few measures to protect their data from it. It seems that people accept privacy risks but still express concerns when prompted (Phelan, Lampe, and Resnick 2016). Interestingly, consumers even seem to seek out more relevant advertising messages that fulfill their specific needs (Kumar and Gupta 2016). More work is needed to understand why consumers like or dislike OBA, specifically because consumers' negative perceptions of OBA are not in line with their behavior to protect themselves or their expectations.

Personalization has an important influence on consumers' responses to OBA. However, studies examining this factor are somewhat limited with respect to the varieties in the levels of personalization. We know that consumers may consider tailored ads to be useful, but they may consider the use of more sensitive information to be creepy and inappropriate. Such feelings can lead to reactance and privacy concerns (e.g., Bleier and Eisenbeiss 2015; Smit, Van Noort, and Voorveld 2014; Ur et al. 2012). However, it is currently not well understood what levels of personalization consumers find acceptable and what they consider creepy. There is ample room for improvement in the understanding of how consumers respond to the usage of various types and amounts of personal data with respect to resistance, acceptance, and the advertising outcomes of OBA. Further research should aim to understand and identify the tipping point, including the point where consumers feel that data collection for OBA becomes too invasive, what they consider acceptable or unacceptable, in what context, and how this affects their responses to different levels of personalization in OBA.

Research suggests it is important to know who is being targeted, as there seem to be individual differences in responses to OBA. Although our framework delineates some important consumer characteristics, theory suggests there are more relevant characteristics to be examined. For instance, the interactive advertising model (Rodgers and Thorson 2000) indicates that consumer responses to interactive advertising depend on their motives for being online and whether those motives are goal oriented. When an ad addresses such motives (e.g., information seeking or entertainment), it is believed to increase the attention, memory, and attitudes toward the ad, which consequently enhance outcomes such as clicks and purchases. This seems particularly applicable to OBA because it is based on online behavior, which is directly connected to online motives.

### Empowering the Consumer

In general, the research suggests that consumers lack relevant knowledge about OBA but do have significant concerns about the collection and use of personal information online. We believe there is both a theoretically and socially relevant gap in our understanding of how we can improve consumers' knowledge and empower them to take actions when they think it is necessary. Our framework shows there are two important gaps with respect to consumer knowledge: There is an absence of research that investigates how OBA characteristics could influence consumer knowledge and abilities and how knowledge would affect OBA outcomes.

More important, further research is needed to gain insights into how we can educate people about OBA and empower them to protect their online privacy. The research has pointed out that consumers might not be able to protect themselves mainly because they do not have the knowledge to assess whether their protective behavior is effective. Using ineffective tools might lead to a false sense of safety. However, consumers are generally positive toward the notion of protecting their privacy (Smit, Van Noort, and Voorveld 2014), such as by clearing their browsing histories or installing ad blockers. In addition, Schaub et al. (2016) found that popular extensions (e.g., Ghostery, DoNotTrackMe, and Disconnect) can increase consumers' privacy awareness.

We thus see an important gap in the literature with respect to the extent to which people engage in self-protection behavior to circumvent websites' online data collection, as well as how we can encourage self-protection. Future research may draw on theoretical models from health communication, such as protection motivation theory (Rogers 1975) and the extended parallel processing model (Witte 1994). These models posit that people's motivations to protect themselves from a specific threat depend on the perceived threat (based on the perceived severity of the threat and one's own susceptibility) and the perceived efficacy of dealing with this threat (based on the efficacy of the response and self-efficacy). When perceived threat and efficacy are high, people are motivated to protect themselves and adapt their behavior.

Although consumers do seem to understand the threat of online data collection and OBA, their efficacy in protecting their online privacy seems to be low. Even when people understand OBA, they still may not be capable of protecting their privacy. Many different tools can be used to mitigate data collection, but not all of these tools are equally effective (Leon, Ur, et al. 2012). Using ineffective tools would therefore cause a false sense of safety. Further research could address this problem and should aim to develop transparency approaches and education that could encourage and help consumers to protect their online privacy. Such research could also help identify the extent to which consumers might need additional (legal) protection.

Another step is to enhance the current approaches of transparency (e.g., informed consent requests, privacy statements, icons). People often do not notice most approaches to achieve transparency, or they ignore them. Therefore, transparency approaches often fail to increase knowledge and awareness of OBA and encourage self-protection. One such example is the OBA icon. Adding some descriptive information to the icon might help consumers understand that an ad is based on their personal online behavior (Leon, Cranshaw, et al. 2012; Van Noort, Smit, and Voorveld 2013). Moreover, transparency can positively influence consumers' perceptions of OBA and ad effectiveness and thus mostly benefits advertisers. More investigation is needed to determine which transparency approach could be effective in truly increasing OBA knowledge.

### Novel Methodological Approaches

Researchers have used various methods to understand the effects and responses to OBA. The most applied method is experimental research, often involving scenario-based approaches (e.g., Jai, Burns, and King 2013; Van Doorn and Hoekstra 2013; Van Noort, Smit, and Voorveld 2013). Field experiments are also common, and some are combined with scenario-based experiments (e.g., Bleier and Eisenbeiss 2015; Sutanto et al. 2013; Tucker 2014). In addition, surveys are often used to examine users' OBA knowledge and perceptions (e.g., Baek and Morimoto 2012; McDonald and Cranor 2010; Turow et al. 2009). Content analysis (Ahrens and Coyle 2011) or qualitative data in the form of interviews (Leon, Ur, et al. 2012; McDonald and Cranor 2010; Ur et al. 2012) and focus groups (Marreiros et al. 2015) are rarely used.

Despite the diversity in methods, there is still much to gain. Examining OBA empirically is a challenge. It is difficult to measure consumers' exposure to OBA and to examine its consequences. The field could thus benefit from methodological innovations. For instance, research is often constrained to scenario-based experiments, as it is often difficult to manipulate behaviorally targeted ads for participants because such ads are based on personal online behavior. Because this information is specific for each individual and difficult for researchers to obtain, it is challenging to develop behaviorally targeted ads in an experimental design. Just like advertisers, a challenge for academic researchers is to determine ways to use big data in the context of OBA and to combine different data sources (Kumar and Gupta 2016). For instance, tracking consumers' browsing behavior could allow researchers to observe how consumers are exposed to OBA and how they respond to it. Such an approach could identify when, for whom, and in what situations OBA is effective.

Furthermore, we believe that research into OBA could benefit from implicit and unobtrusive measures to gain more insights into how consumers process and respond to it. For instance, eye-tracking research into the effects of ads that are personalized based on demographic variables (e.g., name or gender) showed that personalized ads attract more attention than nonpersonalized ads (Bang and Wojdynski 2016). For now, we have no insights into whether behaviorally targeted ads also attract more visual attention. Such insights would be relevant information, as people's attention to ads targeted on the basis of their personal data could explain their responses to the ad.

Finally, OBA research lacks a longitudinal focus, making it difficult to observe developments over several years. Hence, we do not know how consumers' knowledge, perceptions, and responses toward OBA change over time. Because technology develops quickly and consumers' knowledge and attitudes toward OBA might change, research combining panel studies and longitudinal (big) data could offer important new insights.

## IMPLICATIONS

### Implications for Privacy Law in the Area of OBA

Because of OBA's privacy implications, policymakers around the world have taken an interest in OBA (e.g., Article 29 Data Protection Working Party 2010; FTC 2012; Hong Kong Office of the Privacy Commissioner for Personal Data 2014; Office of the Australian Privacy Commissioner 2011). Data privacy laws vary from country to country. However, despite the differences, many national data privacy laws also have common features (Bygrave 2014). Many privacy laws aim to empower the consumer by requiring companies to offer transparency and choices regarding data use. Our review of academic empirical studies shows that consumers understand little about OBA and the related data use, and current transparency approaches are not very effective in increasing understanding. Consumers who do not understand how data are used for OBA cannot make meaningful privacy decisions.

To defend privacy, we propose that policymakers should not merely aim for consumer empowerment but also for protection. Most privacy laws have elements that aim to protect consumers. For instance, many laws require companies to secure the data they collect against data breaches, and in many countries the law has stricter rules for certain types of sensitive data, such as health-related data (EU Data Protection Directive 1995; EU General Data Protection Regulation 2016; OECD 2013). However, to defend privacy, perhaps more and stricter rules are needed for OBA. There may be OBA practices that society should not accept, regardless of whether consumers consent to the practices. Examples include tracking on websites aimed at children and the use of OBA data for online price discrimination.

### Implications for Advertisers

The advertising industry claims that OBA is much more effective than nontargeted ads (Beales 2010; Chen and Stallaert 2014). However, our framework reveals that the effects are more nuanced. OBA can increase click-through rates and purchases, but these effects depend on factors controlled by the advertiser (e.g., information used to personalize the ad and advertiser transparency) and factors that are related to the consumer (e.g., trust in the advertiser, perceived usefulness of the ad, feelings of intrusiveness, privacy concerns). Our framework could help predict the outcomes of OBA.

Advertisers should consider the level of personalization. Ads perceived to be too personal can seem intrusive and lower click-through rates and purchases. In addition, advertisers should be transparent. Overtly informing people about the collection and use of data to personalize ads can benefit the response and outcomes of OBA. Furthermore, OBA may not be beneficial in every situation or audience. OBA seems to have the most positive outcomes for people who are younger, have high levels of online experience, have low levels of privacy concerns, and have narrowly construed preferences.

## OBA AND THE FUTURE

Although our research agenda is not exhaustive, it shows important gaps in the literature and fruitful areas for further work. Personalized and targeted advertising is seen as the future of advertising (Kumar and Gupta 2016; Schultz 2016; Rust 2016). OBA is still far from mature, and it could be seen as an early example of ambient intelligence—technology that senses and anticipates people's behavior to adapt the environment to their inferred needs (Hildebrandt 2010). Currently, behavioral targeting mostly occurs when using computers or smartphones, but the borders between offline and online are fading. Phrases such as the Internet of things, ubiquitous computing (Weiser 1993), and ambient intelligence (Aarts and Wichert 2009) have been used to describe such developments. If objects are connected to the Internet, companies could use the data collected through those objects for OBA. Google (2013) predicts, "A few years from now, we and other companies could be serving ads and other content on refrigerators, car dashboards, thermostats, glasses, and watches, to name just a few possibilities." These developments have important implications for advertisers, consumers, scholars, and public policy, and they open up a whole new field of research.

## ACKNOWLEDGMENTS

This project is supported by the Personalised Communication Project at the University of Amsterdam (http://personalised-communication.net/). The authors would like to thank Prof. Dr. Natali Helberger and Prof. Dr. Claes de Vreese for their valuable comments on an earlier version of this article.